\documentclass{article}

\usepackage{arxiv}

\usepackage[utf8]{inputenc} 
\usepackage[T1]{fontenc}    
\usepackage{hyperref}       
\usepackage{url}            
\usepackage{booktabs}       
\usepackage{amsfonts}       
\usepackage{nicefrac}       
\usepackage{microtype}      
\usepackage{lipsum}
\usepackage{graphics}
\usepackage{epigraph}
\usepackage{graphicx}
\usepackage{siunitx}
\usepackage{boldline}
\usepackage{pdflscape}
\usepackage{calrsfs}
\usepackage{amsmath}
\usepackage{mdframed}
\usepackage{amsfonts,amssymb,amsbsy,tabulary,amsmath}
\usepackage{subfigure}
\usepackage{enumitem}
\usepackage{lscape}
\usepackage{xcolor}
\usepackage{caption}
\title{A comprehensive evaluation of full-reference image quality assessment algorithms on KADID-10k}

\author{
  Domonkos Varga \\
  Department of Networked Systems and Services\\
  Budapest University of Technology\\
}

\begin{document}
\maketitle

\begin{abstract}
Significant progress has been made in the past decade for full-reference image quality assessment (FR-IQA).
However, new large scale image quality databases have been released for evaluating image quality assessment
algorithms. In this study, our goal is to give a comprehensive evaluation of state-of-the-art FR-IQA metrics
using the recently published KADID-10k database which is largest available one at the moment. Our evaluation
results and the associated discussions is very helpful to obtain a clear understanding about the status
of state-of-the-art FR-IQA metrics.
\end{abstract}

\keywords{Image quality assessment}

\section{Introduction}
Image quality assessment (IQA) is an important element of various applications ranging from
display technology to video surveillance and ADAS systems. Furthermore, image quality measurements
require a balanced investigation of visual content and features. Digital images may suffer from
various distortions during transmission, storing, and sharing. Owing to recent developments in
multimedia technology and camera systems, the design of reliable IQA algorithms has attracted considerable
attention. Consequently, IQA has been the focus of many research studies and patents.

In this study, we provide a comprehensive evaluation of 29 full-reference image quality assessment (FR-IQA) metrics
on the recently published KADID-10k database \cite{kadid10k}.

The rest of this study is organized as follows. Subsection \ref{sec:subjective} briefly reviews subjective visual quality assessment.
In Subsection \ref{sec:objective}, the definition and common classification of objective visual quality assessment are given.
In Subsection \ref{sec:eval} the common evaluation of metrics of visual quality assessment algorithms are given.
Section \ref{sec:results} summarizes our evaluation results. Finally, a conclusion is drawn in Section \ref{sec:conc}.

\subsection{Subjective visual quality assessment}
\label{sec:subjective}
Visual signals (digital images and videos) can undergo a wide variety of distortions after their capture during compression,
transmission, and storage. Human observers are the end users of visual content;
thus, the quality of visual signals should ideally be evaluated in subjective user studies in a
laboratory environment involving specialists. During these user studies, subjective quality
scores are collected from each participant. Subsequently, the quality of a visual signal is given
a mean opinion score (MOS), which is calculated as the arithmetic mean of all the individual
quality ratings. In most cases, an absolute category rating is applied, which ranges from 1.0
(bad quality) to 5.0 (excellent quality). Other standardized quality ratings also exist, such as
a continuous scale ranging from 1.0 to 100.0, but Huynh-Thu \textit{et al.} \cite{huynh2011study} noted that there are
no statistical differences between the different scales used for the same visual stimuli.

Several international standards such as ITU BT.500-13 \cite{bt2002methodology}, ITU P910 \cite{itu1999subjective}
have been proposed for performing subjective visual quality 
assessment. As already mentioned, the main goal of subjective visual quality assessment is to assign a score of the user's 
percieved quality to each visual signal in a given set of signals. The resulted assessment might vary
significantly because of many factors such as lightning conditions and the choice of
subjects. 
For visual quality assessment images or videos are displayed for a given period of time and scores can be either qualitatively 
or quantitatively  scaled --- single incentive rating method --- or both test and reference
images can be displayed at the same time --- double incentive rating --- to the observers.
ITU-R BT.500-13 \cite{bt2002methodology} gives detailed recommendations about viewing conditions, monitor
resolution, selection of test materials, observers, test session, grading scales, analysis
and interpretation of the results.
There are four primary methods for the subjective image and video quality rating which are compared 
in \cite{mantiuk2012comparison}.

\begin{itemize}
 \item Single stimulus (SS) or absolute category rating (ACR): test images or videos are displayed on  a screen for  a fixed amount of time, 
after that, they  will disappear from the screen and observers will be asked to rate 
the  quality  of them on  an  abstract  scale  containing  one  of  the  five  categories: 
excellent, good, fair, poor, or bad. All of the test images or videos are displayed randomly. 
In order to avoid quantization artifacts, some methods use continuous rather than 
categorical scales \cite{bt2002methodology}. In Table \ref{table:acr} an example of ACR is shown.
MOS of 5 represents the best image quality, while 1 is the worst image quality.
 \item  Double stimulus categorical rating:  It is similar to the single stimulus method but in  this  method 
both the test and reference signals are being displayed for a fixed amount of time. After 
that, images or videos will disappear from the screen and observers will be asked to rate the 
quality of the test image or video according to the abstract scale described earlier.
 \item  Ordering by force-choice pair-wise comparison: two  images or videos  of  the  same  scene  are  being 
displayed  for observers.  Afterward ,  they  are  asked  to  choose  the  image or the video  with 
higher  quality.  Observers  are  always  required  to  choose  one  image or video
even  if  both images or videos  possess  no  difference. There  is  no  time  limit  for  observers  to
make  the decision. The drawback of this approach is that it requires more trials to compare 
each pair of  conditions.
\item Pair-wise similarity judgments: In this process observers are asked not only 
to choose the image or video with higher quality, but also to indicate the level 
of difference between  them on a continuous scale.
\end{itemize}

\begin{table}[h]
\caption{Example of absolute category rating (ACR) scale. MOS of 5 represents 
the best image quality, while 1 is the worst image quality.}
\centering 
\begin{center}
    \begin{tabular}{ |c|c|c|}
    \hline
 Mean opinion score    &  Perceptual quality & Impairment \\
    \hline
 5 & excellent & Imperceptible\\
 4 & good      & Perceptible but not annoying\\
 3 & fair      & Slightly annoying\\
 2 & poor      & Annoying\\
 1 & bad       & Very annoying\\
 \hline
 \end{tabular}
\end{center}
\label{table:acr}
\end{table}

Different scales can be utilized for the final score, e.g. the percieved quality of a visual signal 
can be calculated as the mean of the scores that each observer assigned to that visual signal 
named Mean Opinion Score (MOS).

Subjective visual quality assessment has some drawbacks which limit their applications:

\begin{itemize}
 \item[-] They are time-consuming and expensive because subjective results are obtained through experiments
 with many observers.
 \item[-] They cannot be part of real-time applications such as image transmission systems.
 \item[-] Their results  depend    on  the  observers'  physical  conditions  and 
      emotional  state.  Factors  such  as  display  device  and lighting condition heavily 
      affect the results of such experiments.
\end{itemize}

Therefore the development of objective visual quality assessment methods that  are  able  to  predict 
the  perceptual  quality  of  visual  signals  is of high importance.

\subsection{Objective visual quality assessment}
\label{sec:objective}
As mentioned in the previous subsection, subjective visual quality assessment is expensive, time consuming, and labor intensive, thereby
preventing its application to real-time systems, such as video surveillance or video streaming.
Moreover, the results obtained by subjective visual quality assessment depend on
the physical condition, emotional state, personality, and culture of the
observers \cite{scott2016personality}.
As a consequence, there is an increasing need for objective visual quality assessment. The classification of
visual quality assessment algorithms
is based on the availability of the original (reference) signal.

\begin{figure}
\centering
\includegraphics[width=0.75\textwidth]{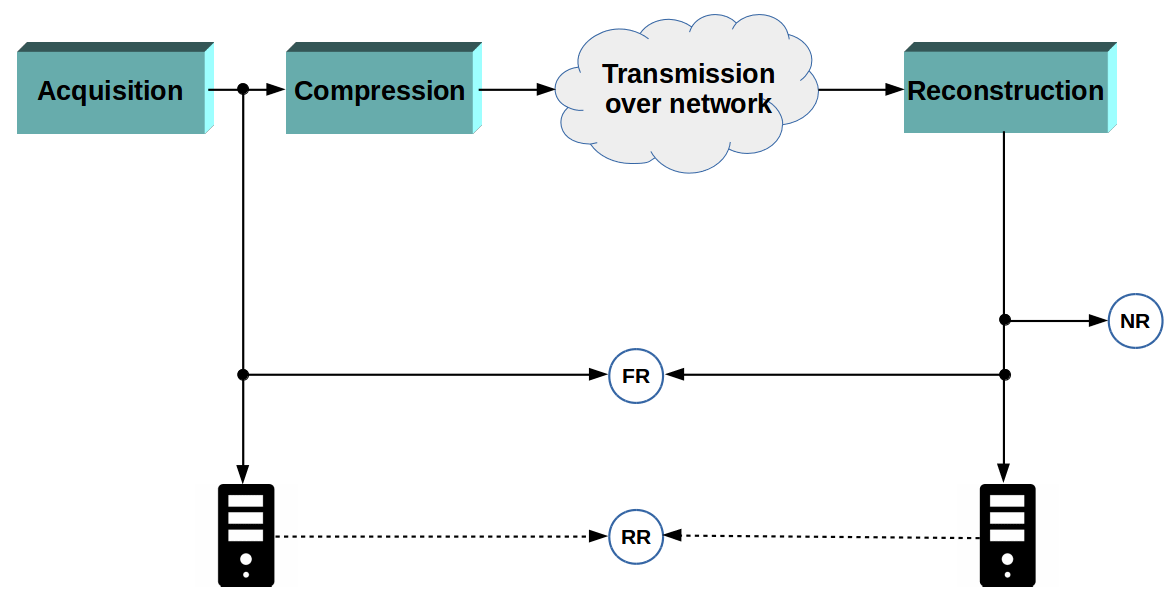}
\caption{Classification of objective visual quality assessment methods: full-reference (FR), reduced-reference (RR), no-reference (NR).}
\label{fig:VisualOverview}
\end{figure}

If a reference signal is not available, a visual quality assessment algorithm is regarded as a no-reference (NR) one.
NR algorithms can be classified into two further groups, where the so-called
distortion-specific NR algorithms assume that a specific distortion is present in the visual
signal, whereas general purpose (or non-distortion specific) algorithms operate on various
distortion types. Reduced-reference (RR) methods retain only part of the information from the reference
signal, whereas full-reference (FR) algorithms have full access to the complete reference
medium to predict the quality scores. Figure \ref{fig:VisualOverview} illustrates the classification
of visual quality assessment methods.

\subsection{Evaluation criteria for assessing visual quality metrics}
\label{sec:eval}
The evaluation of objective visual quality assessment is based on the correlation between the predicted and the ground-truth
quality scores. Pearson's linear correlation coefficient (PLCC) and Spearman's rank order correlation coefficient (SROCC) are
widely applied to this end. The PLCC between data set $A$ and $B$ is defined as
\begin{equation}
    PLCC(A,B) = \frac{\sum_{i=1}^{n} (A_i-\overline{A})(B_i-\overline{B})}{\sqrt{\sum_{i=1}^{n}(A_i-\overline{A})^{2}}\sqrt{\sum_{i=1}^{n}(B_i-\overline{B})^{2}}},
\end{equation}
where $\overline{A}$ and $\overline{B}$ stand for the average of set $A$ and $B$, $A_i$ and $B_i$ denote the $i$th elements of set $A$ and $B$,
respectively. For two ranked
sets \textit{A} and \textit{B} SROCC is defined as
\begin{equation}
    SROCC(A,B)=\frac{\sum_{i=1}^{n} (A_i-\hat{A})(B_i-\hat{B})}{\sqrt{\sum_{i=1}^{n}(A_i-\hat{A})^{2}}\sqrt{\sum_{i=1}^{n}(B_i-\hat{B})^{2}}},
\end{equation}
where $\hat{A} $ and $\hat{B} $ are the middle ranks of set \textit{A} and \textit{B}.
KROCC between dataset \textit{A} and \textit{B} can be calculated as
\begin{equation}
 KROCC(A,B)=\frac{n_c-n_d}{\frac{1}{2}n(n-1)},
\end{equation}
where n is the length of the input vectors, $n_c$ is the number of concordant pairs between \textit{A} and \textit{B}, and
$n_d$ is the number of discordant pairs between \textit{A} and \textit{B}.

\section{Evaluation Results}
\label{sec:results}
As already mentioned, KADID-10k\footnote{Available: http://database.mmsp-kn.de/kadid-10k-database.html} \cite{kadid10k} database
was used to evaluate the performance of the considered 29 FR-IQA metrics
whose original source codes are available. KADID-10k \cite{kadid10k} consists of 81 pristine images and 10,125 distorted images
derived from the pristine images considering 25 different distortion types at 5 intensity levels $(10,125=81\times25\times5)$.
Table \ref{table:noise} shows the considered distortion types and their corresponding numeric codes in KADID-10k \cite{kadid10k}.

As already mentioned, we have evaluated 29 FR-IQA methods on KADID-10k \cite{kadid10k} using their default input parameter
settings (if any). Furthermore, we report on PLCC, SROCC, and KROCC
for the entire database in Table \ref{table:all}.
It can be clearly seen from the results that there is still a lot of space for the improvement of FR-IQA algorithms because none of the considered
state-of-the-art FR-IQA metrics could perform over 0.9 PLCC/SROCC/KROCC. On the whole, HaarPSI \cite{reisenhofer2018haar} and
MDSI \cite{nafchi2016mean} achieved the best results
and significantly outperformed other state-of-the-art methods. Older metrics, such as UQI \cite{wang2002universal}, IFC \cite{sheikh2005information},
QILV \cite{aja2006image}, perform rather weak.
Surprisingly, the performance of the older SSIM \cite{wang2004image} is comparable to more recent FR-IQA methods.
In Tables \ref{table:level_1}, \ref{table:level_2} and \ref{table:level_3}, the PLCC, SROCC, and KROCC values are given with respect to
each distortion level. In KADID-10k \cite{kadid10k}, Level 1 represents the lowest possible distortion level, while Level 5 stands for the highest
distortion level. Surprisingly, some FR algorithms, such as UQI \cite{wang2002universal}, SUMMER\cite{temel2019perceptual}, QILV \cite{aja2006image},
MS-SSIM \cite{wang2003multiscale}, MAD \cite{larson2010most}, IFC \cite{sheikh2005information}, give significantly better results
on images with lower distortion intensity levels than on images with higher distortion intensity levels.
Similarly, Tables \ref{table:noise_1}, \ref{table:noise_2}, and \ref{table:noise_3} summarize the PLCC, SROCC, and KROCC values with respect to
the different distortion types. It can be seen that the FR metrics' performance is not uniform. The measured PLCC/SROCC/KROCC values over different distortion types
may differ very significantly for almost all FR algorithm.
For instance, VSI's \cite{zhang2014vsi} performance over $\#01$ distortion type is among the best, while the performance
over $\#20$ distortion type is rather weak. Moreover, $\#20$ distortion type proves very challenging for all metrics.
The measured data provided in Tables \ref{table:noise_1}, \ref{table:noise_2}, and \ref{table:noise_3} is very useful to improve existing
FR metrics because challenging distortion types can be identified easily. 

\begin{table}[h]
\caption{
Distortion types in
KADID-10k \cite{kadid10k}.
} 
\centering 
\begin{center}
\begin{tabular}{|c | c |}
\hline
Code & Distortion type    \\
\hline
\#01        &  Gaussian blur \\
\#02        &  Lens blur \\
\#03        &  Motion blur \\
\#04        &  Color diffusion \\
\#05        &  Color shift \\
\#06        &  Color quantization \\
\#07        &  Color saturation 1. \\
\#08        &  Color saturation 2. \\
\#09        &  JPEG2000 \\
\#10        &  JPEG \\
\#11        &  White noise \\
\#12        &  White noise in color component \\
\#13        &  Impulse noise \\
\#14        &  Multiplicative noise \\
\#15        &  Denoise \\
\#16        &  Brighten \\
\#17        &  Darken \\
\#18        &  Mean shift \\
\#19        &  Jitter \\
\#20        &  Non-eccentricity patch \\
\#21        &  Pixelate \\
\#22        &  Quantization \\
\#23        &  Color block \\
\#24        &  High sharpen \\
\#25        &  Contrast change \\
\lasthline
\end{tabular}  
\end{center}
\label{table:noise}
\end{table}

\begin{table}[h]
\caption{
Performance comparison on
KADID-10k \cite{kadid10k}.
The best results are typed by \textbf{bold}.
} 
\centering 
\begin{center}
  \begin{tabular}{|c |c| c | c | c|}
\hline
Method & Year&PLCC    & SROCC  & KROCC\\
\hline
BLeSS-FSIM \cite{temel2016bless}         &2016 &0.692 & 0.816 & 0.624 \\
BLeSS-FSIMc \cite{temel2016bless}        &2016 &0.714 & 0.848 & 0.658 \\
BLeSS-SR-SIM \cite{temel2016bless}       &2016 &0.652 & 0.824 & 0.633 \\
CVSS \cite{jia2018contrast}              &2018 &0.804 & 0.861 & 0.678 \\
DSS \cite{balanov2015image}              &2015 &0.791 & 0.860 & 0.674 \\
ESSIM \cite{zhang2013edge}               &2013 &0.551 & 0.823 & 0.634 \\
FSIM \cite{zhang2011fsim}                &2011 &0.677 & 0.829 & 0.639 \\
\end{tabular}  
\end{center}
\label{table:all}
\end{table}

\begin{table}[h]
\caption{
PLCC values of the considered FR-IQA metrics for each distortion level of
KADID-10k \cite{kadid10k} separately. The best results are typed by \textbf{bold} with respect to each distortion level.
Level 1 corresponds to the lowest distortion level, while Level 5 represents the highest distortion level.
} 
\centering 
\begin{center}
  \begin{tabular}{|c | c | c | c| c| c|}
\hline
Method & Level 1    & Level 2  & Level 3 & Level 4 & Level 5\\
\hline
BLeSS-FSIM \cite{temel2016bless}         & 0.557 & 0.585 & 0.609 & 0.560 & 0.540\\
BLeSS-FSIMc \cite{temel2016bless}        & 0.545	& 0.608	&0.647&	0.603 &	0.581\\
BLeSS-SR-SIM \cite{temel2016bless}       & 0.539 & 0.566 & 0.583 & 0.526 & 0.518\\
CVSS \cite{jia2018contrast}              & \textbf{0.665}&	0.726&	0.730&	0.685&	0.626\\
DSS \cite{balanov2015image}              &  0.658&	0.696&	0.699&	0.676&	0.655\\
ESSIM \cite{zhang2013edge}               &  0.432&	0.462&	0.501&	0.474&	0.461\\
FSIM \cite{zhang2011fsim}                &  0.472&	0.530&	0.586&	0.567&	0.567\\
\end{tabular}  
\end{center}
\label{table:level_1}
\end{table}

\begin{table}[h]
\caption{
SROCC values of the considered FR-IQA metrics for each distortion level of
KADID-10k \cite{kadid10k} separately. The best results are typed by \textbf{bold} with respect to each distortion level.
Level 1 corresponds to the lowest distortion level, while Level 5 represents the highest distortion level.
} 
\centering 
\begin{center}
  \begin{tabular}{|c | c | c | c| c| c|}
\hline
Method & Level 1    & Level 2  & Level 3 & Level 4 & Level 5\\
\hline
BLeSS-FSIM \cite{temel2016bless}         & 0.668 & 0.706 & 0.698 & 0.663 & 0.644\\
BLeSS-FSIMc \cite{temel2016bless}        & 0.690 & 0.757 & 0.760 & 0.719 & 0.687\\
BLeSS-SR-SIM \cite{temel2016bless}       & 0.654 & 0.705 & 0.709 & 0.691 & 0.694\\
CVSS \cite{jia2018contrast}               &  0.743&	0.772&	0.767&	0.686&	0.654\\
DSS \cite{balanov2015image}                & \textbf{0.747}&	0.768&	0.761&	0.704&	0.666\\
ESSIM \cite{zhang2013edge}              & 0.664&	0.673&	0.699&	0.712&	0.708\\
FSIM \cite{zhang2011fsim}               &  0.683&	0.704&	0.721&	0.698&	0.690\\
\end{tabular}  
\end{center}
\label{table:level_2}
\end{table}

\begin{table}[h]
\caption{
KROCC values of the considered FR-IQA metrics for each distortion level of
KADID-10k \cite{kadid10k} separately. The best results are typed by \textbf{bold} with respect to each distortion level.
Level 1 corresponds to the lowest distortion level, while Level 5 represents the highest distortion level.
} 
\centering 
\begin{center}
  \begin{tabular}{|c | c | c | c| c| c|}
\hline
Method & Level 1    & Level 2  & Level 3 & Level 4 & Level 5\\
\hline
BLeSS-FSIM \cite{temel2016bless}         & 0.482 & 0.509 & 0.511 & 0.485 & 0.462\\
BLeSS-FSIMc \cite{temel2016bless}        & 0.498 & 0.552 & 0.568 & 0.535 & 0.501\\
BLeSS-SR-SIM \cite{temel2016bless}       & 0.469 & 0.510 & 0.522 & 0.513 & 0.506\\
CVSS \cite{jia2018contrast}              & \textbf{0.549}&	0.581&	0.582&	0.508&	0.477\\
DSS \cite{balanov2015image}              &  0.548&	0.574&	0.571&	0.519&	0.483\\
ESSIM \cite{zhang2013edge}               &  0.475&	0.484&	0.516&	0.531&	0.519\\
FSIM \cite{zhang2011fsim}                &  0.489&	0.508&	0.534&	0.518&	0.501\\
\end{tabular}  
\end{center}
\label{table:level_3}
\end{table}

\begin{landscape}

\begin{table}[h]
\caption{
PLCC values of the considered FR-IQA metrics for each distortion type of
KADID-10k \cite{kadid10k} separately. The best results are typed by \textbf{bold} with respect to each distortion type.
} 
\centering 
\begin{center}
  \begin{tabular}{|c|c|c|c|c|c|c|c|c|c|c|c|c|c|c|c|c|c|c|c|c|c|c|c|c|c|}
\hline
\tiny{Method} & \tiny{\#01}    & \tiny{\#02}  & \tiny{\#03} & \tiny{\#04} & \tiny{\#05} & \tiny{\#06} & \tiny{\#07} & \tiny{\#08} & \tiny{\#09} & \tiny{\#10} & \tiny{\#11} & \tiny{\#12} & \tiny{\#13} & \tiny{\#14} & \tiny{\#15} & \tiny{\#16} & \tiny{\#17} & \tiny{\#18} & \tiny{\#19} & \tiny{\#20} & \tiny{\#21} & \tiny{\#22} & \tiny{\#23} & \tiny{\#24} & \tiny{\#25} \\
\hline
\tiny{BLeSS-FSIM \cite{temel2016bless}} & \tiny{0.782}	&\tiny{0.781}	&\tiny{0.793}	&\tiny{0.745}	&\tiny{0.679}	&\tiny{0.788}	&\tiny{0.461}	&\tiny{0.781}	&\tiny{0.906}	&\tiny{0.928}	&\tiny{0.647}	&\tiny{0.725}	&\tiny{0.662}	&\tiny{0.721}	&\tiny{0.830}	&\tiny{0.882}	&\tiny{0.903}	&\tiny{0.542}	&\tiny{0.834}	&\tiny{0.305}	&\tiny{0.542}	&\tiny{0.704}	&\tiny{0.336}	&\tiny{0.849}	&\tiny{0.609}\\
\tiny{BLeSS-FSIMc \cite{temel2016bless}}& \tiny{0.780}	&\tiny{0.779}	&\tiny{0.791}	&\tiny{0.783}	&\tiny{0.675}	&\tiny{0.857}	&\tiny{0.590}	&\tiny{0.823}	&\tiny{0.908}	&\tiny{0.935}	&\tiny{0.671}	&\tiny{0.747}	&\tiny{0.685}	&\tiny{0.737}	&\tiny{0.846}	&\tiny{0.903}	&\tiny{0.915}	&\tiny{0.602}	&\tiny{0.834}	&\tiny{0.385}	&\tiny{0.577}	&\tiny{0.793}	&\tiny{0.464}	&\tiny{0.847}	&\tiny{0.711}\\
\tiny{BLeSS-SR-SIM \cite{temel2016bless}}&\tiny{0.784}	&\tiny{0.802}	&\tiny{0.825}	&\tiny{0.653}	&\tiny{0.744}	&\tiny{0.736}	&\tiny{0.441}	&\tiny{0.737}	&\tiny{0.901}	&\tiny{0.909}	&\tiny{0.775}	&\tiny{0.826}	&\tiny{0.754}	&\tiny{0.768}	&\tiny{0.833}	&\tiny{0.873}	&\tiny{0.875}	&\tiny{0.489}	&\tiny{0.859}	&\tiny{0.359}	&\tiny{0.591}	&\tiny{0.569}	&\tiny{0.308}	&\tiny{0.844}	&\tiny{0.560}\\
\end{tabular}  
\end{center}
\label{table:noise_1}
\end{table}

\end{landscape}

\begin{landscape}

\begin{table}[h]
\caption{
SROCC values of the considered FR-IQA metrics for each distortion type of
KADID-10k \cite{kadid10k} separately. The best results are typed by \textbf{bold} with respect to each distortion type.
} 
\centering 
\begin{center}
  \begin{tabular}{|c|c|c|c|c|c|c|c|c|c|c|c|c|c|c|c|c|c|c|c|c|c|c|c|c|c|}
\hline
\tiny{Method} & \tiny{\#01}    & \tiny{\#02}  & \tiny{\#03} & \tiny{\#04} & \tiny{\#05} & \tiny{\#06} & \tiny{\#07} & \tiny{\#08} & \tiny{\#09} & \tiny{\#10} & \tiny{\#11} & \tiny{\#12} & \tiny{\#13} & \tiny{\#14} & \tiny{\#15} & \tiny{\#16} & \tiny{\#17} & \tiny{\#18} & \tiny{\#19} & \tiny{\#20} & \tiny{\#21} & \tiny{\#22} & \tiny{\#23} & \tiny{\#24} & \tiny{\#25} \\
\hline
\tiny{BLeSS-FSIM \cite{temel2016bless}}      &\tiny{0.952}          &\tiny{0.900}          &\tiny{0.946}          &\tiny{0.863}          &\tiny{0.694}          &\tiny{0.806}          &\tiny{0.475}          &\tiny{0.863}          &\tiny{0.945}          &\tiny{0.897}          &\tiny{0.820}          &\tiny{0.879}          &\tiny{0.853}          &\tiny{0.891}          &\tiny{0.886}          &\tiny{0.884}          &\tiny{0.866}          &\tiny{0.661}          &\tiny{0.927}          &\tiny{0.400}          &\tiny{0.456}          &\tiny{0.776}          &\tiny{0.361}          &\tiny{0.885}          &\tiny{0.689}\\
\tiny{BLeSS-FSIMc \cite{temel2016bless}}     &\tiny{0.952}          &\tiny{0.900}          &\tiny{0.947}          &\tiny{0.909}          &\tiny{0.683}          &\tiny{0.867}          &\tiny{0.596}          &\tiny{0.903}          &\tiny{0.950}          &\tiny{0.911}          &\tiny{0.835}          &\tiny{0.895}          &\tiny{0.866}          &\tiny{0.911}          &\tiny{0.903}          &\tiny{0.910}          &\tiny{0.889}          &\tiny{0.685}          &\tiny{0.939}          &\tiny{0.429}          &\tiny{0.472}          &\tiny{0.808}          &\tiny{0.481}          &\tiny{0.886}          &\tiny{0.759}\\
\end{tabular}  
\end{center}
\label{table:noise_2}
\end{table}

\end{landscape}

\begin{landscape}

\begin{table}[h]
\caption{
KROCC values of the considered FR-IQA metrics for each distortion type of
KADID-10k \cite{kadid10k} separately. The best results are typed by \textbf{bold} with respect to each distortion type.
} 
\centering 
\begin{center}
  \begin{tabular}{|c|c|c|c|c|c|c|c|c|c|c|c|c|c|c|c|c|c|c|c|c|c|c|c|c|c|}
\hline
\tiny{Method} & \tiny{\#01}    & \tiny{\#02}  & \tiny{\#03} & \tiny{\#04} & \tiny{\#05} & \tiny{\#06} & \tiny{\#07} & \tiny{\#08} & \tiny{\#09} & \tiny{\#10} & \tiny{\#11} & \tiny{\#12} & \tiny{\#13} & \tiny{\#14} & \tiny{\#15} & \tiny{\#16} & \tiny{\#17} & \tiny{\#18} & \tiny{\#19} & \tiny{\#20} & \tiny{\#21} & \tiny{\#22} & \tiny{\#23} & \tiny{\#24} & \tiny{\#25} \\
\hline
\tiny{BLeSS-FSIM \cite{temel2016bless}}  &\tiny{0.814}         &\tiny{0.730}         &\tiny{0.802}         &\tiny{0.675}         &\tiny{0.512}         &\tiny{0.619}         &\tiny{0.331}         &\tiny{0.680}         &\tiny{0.804}         &\tiny{0.728}         &\tiny{0.635}         &\tiny{0.699}         &\tiny{0.667}         &\tiny{0.715}         &\tiny{0.703}         &\tiny{0.694}         &\tiny{0.683}         &\tiny{0.478}         &\tiny{0.759}         &\tiny{0.281}         &\tiny{0.319}         &\tiny{0.581}&\tiny{0.248}&\tiny{0.695}&\tiny{0.486}\\
\tiny{BLeSS-FSIMc \cite{temel2016bless}} &\tiny{0.815}         &\tiny{0.730}	     &\tiny{0.803}         &\tiny{0.738}         &\tiny{0.503}         &\tiny{0.682}         &\tiny{0.425}         &\tiny{0.731}         &\tiny{0.810}         &\tiny{0.753}         &\tiny{0.650}         &\tiny{0.720}         &\tiny{0.681}         &\tiny{0.737}         &\tiny{0.727}         &\tiny{0.727}         &\tiny{0.714}         &\tiny{0.498}         &\tiny{0.779}         &\tiny{0.300}         &\tiny{0.330}         &\tiny{0.613}&\tiny{0.333}&\tiny{0.696}&\tiny{0.549}\\
\end{tabular}  
\end{center}
\label{table:noise_3}
\end{table}

\end{landscape}

\section{Conclusion}
\label{sec:conc}
In this study, we extensively evaluated 29 state-of-the-art FR-IQA methods on KADID-10k \cite{kadid10k} dataset
which is the largest publicly available image quality database containing 81 pristine images and 10,125 distorted ones.
The considered FR-IQA algorithms' prediction performance were reported with respect to the entire database, different distortion intensity levels,
and different distortion types.
\bibliographystyle{unsrt}  
\bibliography{references}  

\begin{thebibliography}{10}

\bibitem{kadid10k}
Hanhe Lin, Vlad Hosu, and Dietmar Saupe.
\newblock Kadid-10k: A large-scale artificially distorted iqa database.
\newblock In {\em 2019 Tenth International Conference on Quality of Multimedia
  Experience (QoMEX)}, pages 1--3. IEEE, 2019.

\bibitem{huynh2011study}
Quan Huynh-Thu, Marie-Neige Garcia, Filippo Speranza, Philip Corriveau, and
  Alexander Raake.
\newblock Study of rating scales for subjective quality assessment of
  high-definition video.
\newblock {\em IEEE Transactions on Broadcasting}, 57(1):1--14, 2011.

\bibitem{bt2002methodology}
RECOMMENDATION ITU-R BT.
\newblock Methodology for the subjective assessment of the quality of
  television pictures.
\newblock 2002.

\bibitem{itu1999subjective}
P~ITU-T~RECOMMENDATION.
\newblock Subjective video quality assessment methods for multimedia
  applications.
\newblock 1999.

\bibitem{mantiuk2012comparison}
Rafa{\l}~K Mantiuk, Anna Tomaszewska, and Rados{\l}aw Mantiuk.
\newblock Comparison of four subjective methods for image quality assessment.
\newblock In {\em Computer graphics forum}, volume~31, pages 2478--2491. Wiley
  Online Library, 2012.

\bibitem{scott2016personality}
Michael~James Scott, Sharath~Chandra Guntuku, Weisi Lin, and Gheorghita Ghinea.
\newblock Do personality and culture influence perceived video quality and
  enjoyment?
\newblock {\em IEEE Transactions on Multimedia}, 18(9):1796--1807, 2016.

\bibitem{reisenhofer2018haar}
Rafael Reisenhofer, Sebastian Bosse, Gitta Kutyniok, and Thomas Wiegand.
\newblock A haar wavelet-based perceptual similarity index for image quality
  assessment.
\newblock {\em Signal Processing: Image Communication}, 61:33--43, 2018.

\bibitem{nafchi2016mean}
Hossein~Ziaei Nafchi, Atena Shahkolaei, Rachid Hedjam, and Mohamed Cheriet.
\newblock Mean deviation similarity index: Efficient and reliable
  full-reference image quality evaluator.
\newblock {\em IEEE Access}, 4:5579--5590, 2016.

\bibitem{wang2002universal}
Zhou Wang and Alan~C Bovik.
\newblock A universal image quality index.
\newblock {\em IEEE signal processing letters}, 9(3):81--84, 2002.

\bibitem{sheikh2005information}
Hamid~R Sheikh, Alan~C Bovik, and Gustavo De~Veciana.
\newblock An information fidelity criterion for image quality assessment using
  natural scene statistics.
\newblock {\em IEEE Transactions on image processing}, 14(12):2117--2128, 2005.

\bibitem{aja2006image}
Santiago Aja-Fernandez, Raul San~Jose Estepar, Carlos Alberola-Lopez, and
  Carl-Fredrik Westin.
\newblock Image quality assessment based on local variance.
\newblock In {\em 2006 International Conference of the IEEE Engineering in
  Medicine and Biology Society}, pages 4815--4818. IEEE, 2006.

\bibitem{wang2004image}
Zhou Wang, Alan~C Bovik, Hamid~R Sheikh, Eero~P Simoncelli, et~al.
\newblock Image quality assessment: from error visibility to structural
  similarity.
\newblock {\em IEEE transactions on image processing}, 13(4):600--612, 2004.

\bibitem{temel2019perceptual}
Dogancan Temel and Ghassan AlRegib.
\newblock Perceptual image quality assessment through spectral analysis of
  error representations.
\newblock {\em Signal Processing: Image Communication}, 70:37--46, 2019.

\bibitem{wang2003multiscale}
Zhou Wang, Eero~P Simoncelli, and Alan~C Bovik.
\newblock Multiscale structural similarity for image quality assessment.
\newblock In {\em The Thrity-Seventh Asilomar Conference on Signals, Systems \&
  Computers, 2003}, volume~2, pages 1398--1402. Ieee, 2003.

\bibitem{larson2010most}
Eric~Cooper Larson and Damon~Michael Chandler.
\newblock Most apparent distortion: full-reference image quality assessment and
  the role of strategy.
\newblock {\em Journal of Electronic Imaging}, 19(1):011006, 2010.

\bibitem{zhang2014vsi}
Lin Zhang, Ying Shen, and Hongyu Li.
\newblock Vsi: A visual saliency-induced index for perceptual image quality
  assessment.
\newblock {\em IEEE Transactions on Image Processing}, 23(10):4270--4281, 2014.

\bibitem{temel2016bless}
Dogancan Temel and Ghassan AlRegib.
\newblock Bless: Bio-inspired low-level spatiochromatic similarity assisted
  image quality assessment.
\newblock In {\em 2016 IEEE International Conference on Multimedia and Expo
  (ICME)}, pages 1--6. IEEE, 2016.

\bibitem{jia2018contrast}
Huizhen Jia, Lu~Zhang, and Tonghan Wang.
\newblock Contrast and visual saliency similarity-induced index for assessing
  image quality.
\newblock {\em IEEE Access}, 6:65885--65893, 2018.

\bibitem{balanov2015image}
Amnon Balanov, Arik Schwartz, Yair Moshe, and Nimrod Peleg.
\newblock Image quality assessment based on dct subband similarity.
\newblock In {\em 2015 IEEE International Conference on Image Processing
  (ICIP)}, pages 2105--2109. IEEE, 2015.

\bibitem{zhang2013edge}
Xuande Zhang, Xiangchu Feng, Weiwei Wang, and Wufeng Xue.
\newblock Edge strength similarity for image quality assessment.
\newblock {\em IEEE Signal processing letters}, 20(4):319--322, 2013.

\bibitem{zhang2011fsim}
Lin Zhang, Lei Zhang, Xuanqin Mou, and David Zhang.
\newblock Fsim: A feature similarity index for image quality assessment.
\newblock {\em IEEE transactions on Image Processing}, 20(8):2378--2386, 2011.

\bibitem{xue2013gradient}
Wufeng Xue, Lei Zhang, Xuanqin Mou, and Alan~C Bovik.
\newblock Gradient magnitude similarity deviation: A highly efficient
  perceptual image quality index.
\newblock {\em IEEE Transactions on Image Processing}, 23(2):684--695, 2013.

\bibitem{liu2011image}
Anmin Liu, Weisi Lin, and Manish Narwaria.
\newblock Image quality assessment based on gradient similarity.
\newblock {\em IEEE Transactions on Image Processing}, 21(4):1500--1512, 2011.

\bibitem{chang2015perceptual}
Hua-wen Chang, Qiu-wen Zhang, Qing-gang Wu, and Yong Gan.
\newblock Perceptual image quality assessment by independent feature detector.
\newblock {\em Neurocomputing}, 151:1142--1152, 2015.

\bibitem{wang2016multiscale}
Tonghan Wang, Lu~Zhang, Huizhen Jia, Baosheng Li, and Huazhong Shu.
\newblock Multiscale contrast similarity deviation: An effective and efficient
  index for perceptual image quality assessment.
\newblock {\em Signal Processing: Image Communication}, 45:1--9, 2016.

\bibitem{temel2015persim}
Dogancan Temel and Ghassan AlRegib.
\newblock Persim: Multi-resolution image quality assessment in the perceptually
  uniform color domain.
\newblock In {\em 2015 IEEE International Conference on Image Processing
  (ICIP)}, pages 1682--1686. IEEE, 2015.

\bibitem{kolaman2011quaternion}
Amir Kolaman and Orly Yadid-Pecht.
\newblock Quaternion structural similarity: a new quality index for color
  images.
\newblock {\em IEEE Transactions on Image Processing}, 21(4):1526--1536, 2011.

\bibitem{zhang2010rfsim}
Lin Zhang, Lei Zhang, and Xuanqin Mou.
\newblock Rfsim: A feature based image quality assessment metric using riesz
  transforms.
\newblock In {\em 2010 IEEE International Conference on Image Processing},
  pages 321--324. IEEE, 2010.

\bibitem{yang2018rvsim}
Guangyi Yang, Deshi Li, Fan Lu, Yue Liao, and Wen Yang.
\newblock Rvsim: a feature similarity method for full-reference image quality
  assessment.
\newblock {\em EURASIP Journal on Image and Video Processing}, 2018(1):6, 2018.

\bibitem{zhang2012sr}
Lin Zhang and Hongyu Li.
\newblock Sr-sim: A fast and high performance iqa index based on spectral
  residual.
\newblock In {\em 2012 19th IEEE international conference on image processing},
  pages 1473--1476. IEEE, 2012.

\bibitem{sheikh2004image}
Hamid~R Sheikh and Alan~C Bovik.
\newblock Image information and visual quality.
\newblock In {\em 2004 IEEE International Conference on Acoustics, Speech, and
  Signal Processing}, volume~3, pages iii--709. IEEE, 2004.

\end{thebibliography}






\end{document}